\newif\ifpdf
\ifx\pdfoutput\undefined
\pdffalse 
\else
\pdfoutput=1 
\pdftrue \fi

\newif\iffinal
\finaltrue

\documentclass[11pt]{article}
\usepackage[notref,notcite]{showkeys}
\usepackage{amsmath, amsthm}
\usepackage{amsmath}
\usepackage{amsfonts}
\usepackage{amssymb}
\usepackage{verbatim}
\usepackage{graphicx}
\usepackage{graphics}
\usepackage{amscd}
\usepackage{amsthm}

\newtheoremstyle{thm}{1.5ex}{1.5ex}{\itshape\rmfamily}{}
{\bfseries\rmfamily}{}{2ex}{}

\newtheoremstyle{rem}{1.3ex}{1.3ex}{\rmfamily}{} 
{\itshape}
{} {1.5ex}{}

\newtheorem{thm}{Theorem}[section]

\newtheorem{cor}[thm]{Corollary}
\newtheorem{lemma}[thm]{Lemma}
\newtheorem{prop}[thm]{Proposition}

\theoremstyle{definition}

\newtheorem{defn}[thm]{Definition}

\numberwithin{equation}{section}

\newcommand{\mbbT}{\mathbb T}
\newcommand{\mbfE}{\mathbf E}
\newcommand{\sfS}{\sf S}
\newcommand{\sfA}{\sf A}
\newcommand{\sfE}{\sf E}
\newcommand{\mcH}{\mathcal H}
\newcommand{\mcK}{\mathcal K}
\newcommand{\mcS}{\mathcal S}
\newcommand{\mcA}{\mathcal A}
\newcommand{\mcE}{\mathcal E}
\newcommand{\mcB}{\mathcal{B}}

\begin{document}

\title
{\Large{Random Cluster Models on the Triangular Lattice}}

\author
{\large{L. Chayes$^1$and H. K.  Lei$^1$}}
\maketitle

\vspace{-4mm}
\centerline{${}^1$\textit{Department of Mathematics, UCLA, Los Angeles,
California, USA}}

\begin{quote}
{\footnotesize {\bf Abstract:}
We study percolation and the random cluster model on the triangular lattice with 3-body interactions.  Starting with percolation, we generalize the star--triangle transformation: We introduce a new parameter (the 3-body term) and identify configurations on the triangles solely by their connectivity.  In this new setup, necessary and sufficient conditions are found for positive correlations and this is used to establish regions of percolation and non-percolation.  Next we apply this set of ideas to the $q>1$ random cluster model:  We derive duality relations for the suitable random cluster measures, prove necessary and sufficient conditions for them to have positive correlations, and finally prove some rigorous theorems concerning phase transitions.

} \footnotesize
\end{quote}

\section{\large{Introduction}}
The study of duality relations for $2D$-Potts systems is not a new 
topic.   Indeed, it is older than the model itself in the sense that 
\cite{KW} and \cite{AT} provided special cases long before the general Potts 
spin--systems were formulated.  While we will not dwell on the historical aspects of this subject, it is worth remarking that this line of study has had immeasurable impact on the entirety of two--dimensional statistical mechanics.  
Notwithstanding, the usual derivations of duality for 
Potts models (see \cite{Wu-rev} and references 
therein) suffer in three respects which we will describe in increasing 
order of importance:

$\bullet$ (I)  There are informal aspects to many of the derivations 
and thus some effort -- presumably small -- would be needed to elevate 
these derivations to the status of mathematical theorems.

$\bullet$ (II)  The various standard techniques, which include mapping 
to vertex models or the introduction of dual--spin variables in the 
form of constraints, do not include all relevant values of parameters.  
In particular, the dual--constraints approach only makes sense for 
integer $q \geq 2$.  It is only as an afterthought that duality relations for 
continuous $q$'s are inferred from the analytic structure of the 
formulas produced for the integer $q$'s.

While we do not necessarily regard these two issues as being of great 
urgency, the third issue is considered to be pertinent both by 
mathematicians and  physicists.

$\bullet$ (III)  The result of a typical duality relation is the 
identification of the  free energies at dual parameter values. 
Hence, as concerns the subject of \emph{phase transitions} one is always left with an unsatisfactory provisional statement: 
{\em If} there is but a single non--analyticity, {\em then} this must 
occur at the self--dual point.

It should be remarked that this third issue is certainly not 
``academic''.  In particular, in the so called $rs$--models \cite{DR}, which are 
generalizations of the Ashkin--Teller and/or the $q$--state Potts models (with $q = r \times s$) 
there is a self--dual line through an intermediate phase where, 
apparently, nothing of interest transpires; c.f. \cite{Weg}, \cite{Pf} and \cite{CM1}.  

As an alternative to the ``usual methods'', it is possible to establish duality via \emph{graphical representations}, in particular the FK--representation \cite{FK}, whereby the duality shows up on the level of the representation itself.  Duality in this context is akin to (and a generalization of) the elementary sort of duality found in Bernoulli percolation.  Hence, using percolation based techniques, genuine irrefutable statements can be made concerning the  presence of phase transitions at points of self--duality.  For example, on the square lattice, duality of the random cluster models has been used to establish rather sharp theorems concerning their phase structures \cite{BC}, \cite{CS}.

In this work we will study the $q$--state Potts models -- and their associated random cluster  representations -- on the triangular lattice.  For these problems, the derivation is considerably more intricate than the square lattice; one must first go through the intermediate honeycomb lattice.  The inevitable consequence of this contortion is the production of extra correlations in the dual model.  In the language of spin-systems, these correlations translate into the phrase ``three body interactions'' but we iterate that the phenomenon  is quite general and occurs even for percolation ($q=1$).  Well known exceptions to this rule are (i)  The Ising spin--system at all couplings and (ii) A special point, called the star--triangle point, where by a miracle, the correlations in the dual model vanish.  Since the star--triangle point is also a point of self--duality, it is readily identified as the transition point as was done in \cite{KJ}.  However, to the authors' knowledge, it is only for the case of percolation (\cite{Wierman1}, \cite{Wierman2}) that a rigorous theorem along these lines has been established.  

The perspective of this work is that since we are generically stuck with the additional correlations after duality, then they should be in the model from the outset.  We find that with the additional freedom of ``three body interactions'', duality becomes a straightforward map in a two--dimensional space that has a self--dual {\em curve} of fixed points.  One of the points on this curve -- and of no particular significance -- is the star--triangle point.  This general picture has been known (and under appreciated) for quite some time:  Duality relations on the level of free energies are derived in \cite{Wu-rev} using the methods of \cite{Burk} -- here for integer $q\geq 2$.  Additional results along these lines are obtained in \cite{Wu-Lin}, \cite{Wu-Zia} and \cite{BaxterTA} via relations to vertex models.  A cornerstone of the former work is a graphical expansion akin to what is developed here.  However, in these works the representation was only employed as an auxiliary device.  The full potential for relating percolation phenomena in the graphical representations to phase transitions (as defined by other means) and the use of the interplay between direct and dual representations to elucidate this phenomena was not exploited.

From the perspective of rigorous analysis, a significant problem emerges at the outset.  In particular, the sorts of additional correlations introduced are not necessarily {\em positive} correlations.  
E.g. for the spin--systems, the extra interactions are, as often as not, antiferromagnetic.  While this may or may not alter the nature of the transition, it is an enormous technical obstacle since nearly all probabilistic arguments concerning systems of this sort are based on the positivity of correlations.  To overcome these difficulties, we must introduce a reduced state space for the graphical models wherein positive correlations can be re-established.  Notwithstanding, our techniques do not cover the entirety of the self--dual curve but this could in principle be accomplished by an extension of our scheme.  Further, to avoid technical complications we deal exclusively with the isotropic case whenever possible: \emph{A priori}, all three edges of the triangle have the same probability of being occupied.  One might also, with some effort, extend various results proved here to the anisotropic cases.  

The remainder of this paper is organized as follows: In section two, we examine the case of percolation where the necessity of introducing local correlations is underscored.  Here the star-triangle duality is generalized and relatively complete results for the phase diagram are derived.  In section three, we study this problem for the $q>1$ random cluster models.  The duality of \cite{Wu-Lin} and \cite{BaxterTA} are derived by graphical methods and we characterize the conditions for positive correlations.  Finally, in section four, we show that in the region where correlations are positive, there is a phase transition (or at least critical behavior) at all points of self-duality.

\section{\large{Generalized Star--Triangle Relations -- Percolation}}
\subsection{\large{The Classical Star--Triangle Situation}}
In order to motivate our work, we first briefly describe the classical 
star-triangle relation.  As mentioned above, we will treat the isotropic case, so let $p$ be the probability that a bond is occupied.  
Now on any given triangle there are eight possible configurations; we 
denote their respective probabilities by $e$ (empty), $s$ (single), $d$ 
(double) and $a$ (full).  Thus, for example,
$s$ = $p(1-p)^2$.  Under the usual sort of planar duality, the 
triangular lattice problem becomes a problem on the honeycomb lattice 
where we could also associate a bond probability e.g.  $p^\bigstar = 
1-p$.
Considering only connectivity properties and integrating out the 
central vertex returns us to a problem on the triangular lattice (but 
with the triangles inverted).  Using $e^*$, $s^*$,  $d^*$ and $a^*$ to 
denote probabilities of the corresponding configurations, we easily 
arrive at
\begin{equation}
\label{22a}
e^*  =  p^3 + 3p^2(1-p),
  \end{equation}
  \begin{equation}
\label{22b}
s^*  =   p(1-p)^2,
  \end{equation}
  \begin{equation}
\label{22c}
3d^* + a^*  =  (1-p)^3.
  \end{equation}
Ostensibly, one would like to define a $p^*$ such that the right hand 
sides of (\ref{22a}), (\ref{22b}), and (\ref{22c}) are, respectively, 
$(1-p^*)^3$, $p^*(1-p^*)^2$ and $(p^*)^3 + 3(p^*)^2(1-p^*)$.  However, 
for general $p\in(0,1)$, this cannot be done -- there are just too many 
equations.  Explicitly, if we try to force this sort of duality, this 
in turn forces $p$ to a particular value which, in fact, is the one for 
which $p = p^*$.
To see this, if we substitute (\ref{22a}) into (\ref{22b}) we get, in 
the variables $R = p/(1-p)$ and $R^* = p^*/(1-p^*)$, the equation 
$RR^*(R^* + 3) = 1$.  But the similar procedure on  (\ref{22c}) and 
(\ref{22b})
gets us $RR^*(R + 3) = 1$ thence any non-trivial solution requires $R = R^*$.
At $p^* = p$, we see that $p$ must satisfy:
\begin{equation}
\label{23} 
p^3 - 3p + 1 = 0,
\end{equation}
which is of course the self-dual point of the classic star-triangle relation.  

\subsection{\large{Introduction of Correlations}}
Overall, the above situation is clearly {\em not} suitable for the development of a general theory of duality.  Clearly, if we wish to salvage this situation, the next step would be to put in some sort of correlations.  A manageable way to implement correlations -- which has its analogs in physical systems, c.f. subsection \ref{phisiks} -- is to introduce correlations within triangles but to keep separate triangles independent.    (Here, of course, we refer only to ``up-pointing'' triangles; configurations on the ``down-pointing'' triangles will be determined from the former.)  

A secondary consequence of the above duality experiment (on a single triangle) is the observation that, when the rinse cycle is finished, the dual model does not really distinguish between the double and full configurations.  This is due to the fact that all we track are connectivities between sites and, in both situations, the triangle is fully connected.

In this spirit, we might as well confine all of our attention to the three types of configurations listed in (\ref{22a}), (\ref{22b}), and (\ref{22c}); e.g., we \emph{define} our model to have only five configurations on each triangle, namely empty, three singles and a full.  So (in the fully isotropic case) we have five parameters: $e$, $s$ and $a$ with $e+3s+a = 1$. We state without proof the following proposition concerning this model on the triangular lattice: 
\begin{prop}\label{1}
Consider the model on the triangular lattice in which configurations on the up-pointing triangles are independent and confined to empty, singles and full with respective probabilities $e$, $s$ and $a$.  Then this model is dual to the model with parameters $e^*$, $s^*$ and $a^*$ which are given by 
\begin{equation}
\label{24a}
e^* = a
\end{equation}
\begin{equation}
\label{24b}
s^* = s
\end{equation}
\begin{equation}
\label{24c}
a^* = e
\end{equation}
In particular, the condition for self-duality is just $a = e$.

\end{prop}

We make a simple observation which will be useful in the next subsection:
\begin{cor}For the parameters $a$, $e$ as above and for $r \in [0, 1]$, the curve $a+e=r$ is invariant under the $*$--map.\end{cor}

In order to translate all of this into a statement about percolation properties of the model we will need to establish some FKG-type properties of the system.  Since separate up-pointing triangles are independent this amounts to a problem on a single triangle.  Here, unfortunately, we must prove the result for the anisotropic case as it will be needed later.  First we need some basic definitions.

\begin{defn}
Let $\Omega$ be a probability space with probability measure $P$.  Let $A \subset \Omega$ be an event and let $\omega \in \Omega$.  Then the \emph{indicator function} $\mathbf{1}_A$ is defined by
\begin{equation*}
\mathbf{1}_A(\omega) = \begin{cases}
      & \text{1 \hspace{5mm} if $\omega \in A$}, \\
      & \text{0 \hspace{5 mm} otherwise}.
\end{cases}
\end{equation*}
If $f$ is a function on $\Omega$, then $\mathbf{E}(f)$, the \emph{expectation} (or mean value) of $f$ is defined to be
\begin{equation*}
\mathbf{E}(f) = \int_\Omega f(\omega) dP(\omega).
\end{equation*}
Finally, we say the functions $f$ and $g$ have \emph{positive correlations} if
\begin{equation*}
\mathbf{E}(fg) \geq \mathbf{E}(f) \mathbf{E}(g).
\end{equation*} 
\end{defn}
  
\begin{thm}\label{2t} Consider the above described 5-state system realized as bond configurations on a triangle: Let $[\sfS]_1$, $[\sfS]_2$ and $[\sfS]_3$ denote the events that the three various sides of the triangle are the sole bonds occupied with $[\sfA]$ and $[\sfE]$ denoting the full and empty configurations.  Let $\nu$ denote a measure on this system and let us denote the respective probabilities of the above-mentioned by $s_1$, $s_2$, $s_3$, $a$ and $e$.  It is assumed without loss of generality that $s_1\geq s_2 \geq s_3$.  Then the necessary and sufficient condition for $\nu$ to have positive correlation is
\begin{equation*}
ae\geq s_1(s_2+s_3)
\end{equation*}
\end{thm}

\noindent{\bf Proof: }
To prove the necessity of the condition $ae \geq s_1(s_2 + s_3)$, note that if $f(s_1) = 0$, $f(s_2) = 1=f(s_3) = 1$, $f(a)=1$ and $f(e)=0$, and $g(s_1) = 0$, $g(s_2)=g(s_3) = 1$, $g(a) = 1$ and $g(e)=0$, then $\mathbf{E}(fg) \geq \mathbf{E}(f) \mathbf{E}(g)$ gives exactly that $ae \geq s_1(s_2 + s_3)$. 
For sufficiency, we aim to show that
\begin{equation}
\label{pc}
\mathbf{E}(fg)-\mathbf{E}(f)\mathbf{E}(g)\geq 0
\end{equation}
To simplify matters we first note that (\ref{pc}) is not changed by adding constants to $f$ and $g$.
Thus we may assume that $f$ and $g$ are overall non-negative and (by subtracting $f(E)$ and $g(E)$ respectively) vanish on the lowest configuration.  Similarly, the truth or falsehood of (\ref{pc}) is unaffected by the scaling of $f$ and $g$ so we may as well assume that $f([\sfA]) = g([\sfA]) = 1$.

Next let $\sigma$ be a permutation on three letters such that $f([\sfS]_{\sigma_1}) \geq $$f$$([\sfS]_{\sigma_2}) \geq $$f$$([\sfS]_{\sigma_3})$.  Then we are down to six parameters: for convenience let $x_1$, $x_2$, $x_3$ denote $f([\sfS]_{\sigma_1})$, $f([\sfS]_{\sigma_2})$, and $f([\sfS]_{\sigma_3})$, respectively.  Similarly define $y_1$, $y_2$, and $y_3$ for $g$.  We assume that some of these parameters are non-trivial, for otherwise the theorem is already proved.

Next we observe that any increasing function is automatically positively correlated with $\mathbf{1}_{[\sfA]}$, the indicator of the top configuration.  Indeed (with all of our simplifications enforced),
	$\mathbf{E}(\mathbf{1}_{[\sfA]} g) = a$, whereas $\mathbf{E}(\mathbf{1}_A)\mathbf{E}(g) = a\mathbf{E}(g)$, which is smaller.  Thus, the quantity $\mathbf{E}(fg)-\mathbf{E}(f) \mathbf{E}(g)$ will decrease if we subtract from $f$ the function $\lambda\mathbf{1}_A$ with $\lambda > 0$.  However, in order to keep $f$ increasing, the most we can subtract is $\lambda = 1 - max\{x_1, x_2, x_3\} = 1-x_1$, by assumption.  Thus, after this subtraction and more rescaling, we have that $x_1 = 1$.
	
Similar considerations show that $min\{x_1, x_2, x_3\} = 0$.  To see this one first observe that $g$ is always positively correlated with the function $1 - \mathbf{1}_{[\sfE]}$.  Then subtracting from $f$ the function $x_3(1-\mathbf{1}_{[\sfE]})$ (where by assumption $x_3 = min\{x_1, x_2, x_3\}$) and rescaling again gives the desired conclusion.

Given all these simplifications, we now have $\mathbf{E}(fg) = a + s_{\sigma_1}y_1 + s_{\sigma_2}x_2y_2$ and $\mathbf{E}(f)\mathbf{E}(g) = (a + s_{\sigma_1}+s_{\sigma_2}x_2)(a + s_{\sigma_1}y_1+s_{\sigma_2}y_2+s_{\sigma_3}y_3)$.  Since the goal is to show that $\mathbf{E}(fg) \geq \mathbf{E}(f)\mathbf{E}(g)$, we may assume that $y_1 = 0$ and $y_3 = 1$, since the coefficient of $y_1$ in $\mathbf{E}(f)\mathbf{E}(g)$ is smaller than in $\mathbf{E}(fg)$ and $y_3$ does not even occur in $\mathbf{E}(fg)$.

Next one can check that $f$ is positively correlated with $\mathbf{1}_{[\sfS]_{\sigma_2}}+ \mathbf{1}_{[\sfS]_{\sigma_3}} +\mathbf{1}_{[\sfA]}$: To see this observe that $\mathbf{E}((\mathbf{1}_{[\sfS]_{\sigma_2}}+ \mathbf{1}_{[\sfS]_{\sigma_3}} +\mathbf{1}_{[\sfA]})f)=(x_2s_{\sigma_2}+a)(s_{\sigma_2}+s_{\sigma_3}+a+s_{\sigma_1}+e)$ whereas $\mathbf{E}(f)\mathbf{E}(\mathbf{1}_{[\sfS]_{\sigma_2}}+ \mathbf{1}_{[\sfS]_{\sigma_3}} +\mathbf{1}_{[\sfA]})=(xs_{\sigma_2}+a+s_{\sigma_1})(s_{\sigma_2}+s_{\sigma_3}+a)$, so the difference is $ae-s_{\sigma_1}(s_{\sigma_2}+s_{\sigma_3})$, which is positive by hypothesis.  It is also easy to check that $g-y_2(\mathbf{1}_{S_{\sigma_2}}+ \mathbf{1}_{S_{\sigma_3}} +\mathbf{1}_A )$ is still increasing.  Also, note that if $y_2$ was equal to one before the subtraction, then after the subtraction $g\equiv0$ and again the conclusion of the theorem holds trivially, so we may as well assume $y_2 \neq 1$.  As before, Subtracting and renormalizing, we acquire $y_2=0$, which immediately implies that $x_2=1$ since that maximizes $\mathbf{E}(f)\mathbf{E}(g)$ without changing $\mathbf{E}(fg)$.

To summarize we are down to $f([{\sfE}])=g([{\sfE}])=0$, $f([{\sfS}]_{\sigma_3})=g([{\sfS}]_{\sigma_1})=g([{\sfS}]_{\sigma_2})=0$, $f([{\sfS}]_{\sigma_1})=f([{\sfS}]_{\sigma_2})=g([{\sfS}]_{\sigma_3})=1$, and $f([{\sfA}])=g([{\sfA}])=1$, so for positive correlation we need 
\begin{equation*}a \geq (a+s_{\sigma_1}+s_{\sigma_2})(a+s_{\sigma_3}),\end{equation*} 
which is true if $ae\geq s_{\sigma_3}(s_{\sigma_1}+s_{\sigma_2})$.  The right hand side is clearly maximized when $\sigma_3=1$ (since by assumption $s_1$ is the maximum of $s_1$, $s_2$, and $s_3$), and we obtain $ae\geq s_1(s_2+s_3)$ as claimed.
\qed

\medskip
\noindent {\bf Remark (a).} It is clear that the standard FKG technology does not extend to the present case.  Indeed, if we view our system as $\{0,1\}^3$, but restrict our attention to measures which assigns weight zero to the double edge configurations, then it is obvious that the FKG lattice condition \emph{fails} for any such measure.  On a slightly more subtle level, regarding $\{[\sfA], [\sfS]_1, [\sfS]_2, [\sfS]_3, [\sfE]\}$ as simply a partially ordered set with lattice structure given by $X \vee Y = { inf} \{Z|X\preceq Z$ and $Y\preceq Z\}$ and $X \wedge Y = sup\{Z|Z \preceq X$ and $Z\preceq Y\}$, it is not hard to see that the FKG lattice condition holds whenever $ae \geq s_1s_2$.  This is in apparent contradiction with (the necessity part of) Theorem (\ref{2t}).  However, the connection between the lattice condition and positive correlation hinges on the fact that the lattice satisfies distributivity, which is a property that our lattice lacks, as $[\sfS]_1 = [\sfS]_1 \wedge ([\sfS]_2 \vee [\sfS]_3) \neq ([\sfS]_1 \wedge [\sfS]_2) \vee ([\sfS]_1 \wedge [\sfS]_3) = [\sfE]$.

\smallskip
\noindent{\bf Remark (b).}
We observe that $ae \geq 2s^2$ implies that $a^*e^* \geq 2(s^*)^2$ by (\ref{24a}), (\ref{24b}), and (\ref{24c}), so the $*$-map takes the region of positive correlation into itself.

\smallskip
\noindent{\bf Remark (c).}
It is noted that for independent bonds, at density $p$, the condition $ae \geq 2s^2$ is well-satisfied.  But supposing we write
\begin{equation}
\label{32a}
e = (1-p)^3(1-t),
\end{equation}
\begin{equation}
\label{32b}
s = p(1-p)^2(1-t),
\end{equation}
and
\begin{equation}
\label{32c}
a = (p^3 + 3p^2(1-p))(1-t) + t
\end{equation}
(as we will have occasion to do when we discuss magnetic systems) and again consider, with the obvious interpretations, our old eight configurations. Then it is clear that the correlations between bonds are positive if and only if $t \geq 0$.  However, our condition $ae \geq 2s^2$ is satisfied for values of $t$ which are considerably negative.

\subsection{Phase Diagram}
\begin{thm}\label{PD}
Consider the correlated percolation model on the triangular lattice as defined previously which has parameters $e$, $s$ and $a$; the parameters are described by points in the $ae$--plane.  Supposing that $ae \geq 2s^2$, then in the region

$$
a+e >r_0\equiv\frac{2\sqrt{2}}{3+2\sqrt{2}},
$$ 
 the following hold\\
(1) The region $a>e$ has a (unique) infinite cluster.\\
(2) The region $a<e$ has no infinite cluster and is characterized by exponential decay of correlations.\\
(3) The line $a=e$ has no infinite cluster of either type but power law (lower bounds) on the decay of correlations.\\
These results are summarized in Figure 1.

\newcounter{obrazek}
\begin{figure}[t]
\refstepcounter{obrazek}
\label{fig1}
\vspace{.2in}
\ifpdf
\centerline{\includegraphics[width=4.1in]{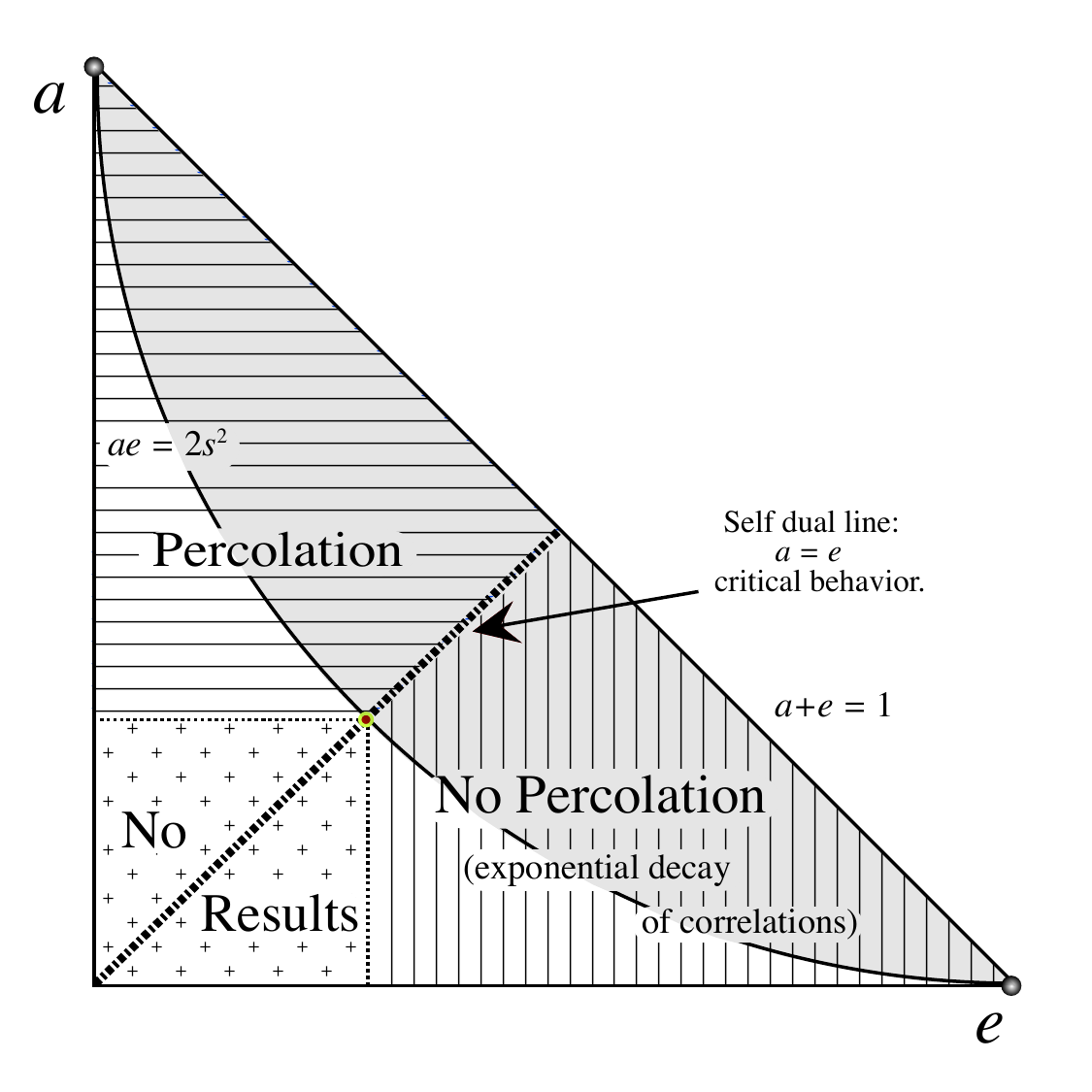}}
\else
\fi
\caption{\footnotesize{
Phase diagram for percolation problem on the triangular lattice; variable $s$ suppressed.  The line $a=e$ is the self--dual line.  The curve $ae=2s^2$ separates the regions with and without positive correlations.  Within the region of positive correlations, $a>e$ is the percolation phase, $a<e$ non--percolating with exponential decay of connectivities and percolation of the \emph{dual} model.  These phases are divided by the self--dual line, where there is no percolation of either type and critical behavior is observed.  Some of these results may be extended out of the region of positive correlations by monotonicity.}}
\end{figure}

\end{thm}
\noindent{\bf Proof} (sketch): We will be brief since the major ingredients are transcriptions with minimal modifications of the well-known results from standard percolation theory.  Our setup will be as follows: we will fix the value of $a+e$, denoting this by $r$, and write $a=\lambda r$, $e=(1-\lambda) r$, $0\leq \lambda \leq 1$.  We will denote by $\lambda_c(r)$ the (purported) threshold above which there is percolation (Notwithstanding, we do not ``yet'' know that there will be percolation even if $\lambda=1$).  Notice by Proposition (\ref{1}) and its corollary that, in these circumstances, the duality takes $\lambda$ to $1-\lambda$.

Our first claim is that the result on the exponential decay of connectivities below threshold applies whenever $r>0$ \cite{M},\cite{MMS}.  
The starting point is the adaptation of Russo's formula \cite{R} to the current situation.  For an increasing event $\mcA$, a triangle $t$ is pivotal if, when empty, the event $\mcA$ does not occur but if fully occupied then it does.  Denoting by $\mathbf{P}_{r, \lambda}$ the probability measure with parameters $a = \lambda r$, etc, and 
$\mathbf{E}_{r, \lambda}$ for the corresponding expectation, the modification of Russo's formula is easily shown to be

\begin{equation*}
\frac{\partial\mathbf{P}(\mcA)}{\partial\lambda} = r \mathbf{E}_{r,\lambda}(|\delta \mcA|),
\end{equation*} 
\noindent
where $|\delta \mcA|$ denotes the number of pivotal triangles for the event $\mcA$.

Next, we denote by $A_n$ the event that the origin is connected by occupied bonds to the boundary of a ``ball'' of radius $n$.  It is clear that the basic ``chain of sausages'' picture holds in this context (with paths of bonds replaced by clusters of triangles) and at the endpoint of each sausage, a pivotal triangle.  We note that for the present setup two events are said to occur \emph{disjointly} if they are determined on the configurations in disjoint sets of triangles. Thus, using the more general Reimer's inequality \cite{Reimer} in place of the van den Berg--Kesten inequality one can follow the standard derivations to obtain
\begin{equation*}
\mathbf{E}_{r,\lambda}(|\delta A_n|) \geq \frac{n}{\sum_{k=1}^{n} \mathbf{P}_{r,\lambda}(A_k)} - 
{\rm const}.
\end{equation*} 

Thereafter, some tedious analysis shows that if at some $\lambda_0$, $\mathbf{P}_{r, \lambda_0} (A_n) \rightarrow 0$ then for all $\lambda < \lambda_0$, $\exists \Psi > 0$ such that $\mathbf{P}_{r, \lambda}(A_n) \leq e^{-\Psi(\lambda, r)n}$; in particular there is exponential decay of connectivities.  However, standard 2D arguments show that once the direct model has rapid decay of correlations, the dual model percolates.  (E.g, if there is no connection between points on the $x$-axis with $x < -L$ and points with $x > +L$ than some dual point with $x$-coordinate in the vicinity of the gap is connected to infinity).  

Using duality this immediately implies that $\lambda_c \leq 1/2$: Any other possibility would imply percolation of the dual model at values of $\lambda$ greater than $1/2$ which, by duality, implies percolation at $\lambda$--values less than $1/2$, contradicting the possibility of any other possibility.

For general values of parameters, the results of \cite{BK} apply which rules out the possibility of multiple infinite clusters (of the same type).  In the region of positive correlation ($r\geq r_0$) the results of \cite{Russo} and \cite{GGR} (see also the proof by Zhang, 1988, unpublished) demonstrates that infinite clusters of the opposite type cannot coexist.  This implies that there cannot be percolation of either type on the self-dual line, i.e. that $\lambda_c \geq 1/2$ so that $\lambda_c = 1/2$.

Finally, to prove power law lower bounds on the decay of correlations, we observe that for appropriate rectangles of length--scale $L$, there is either a left--right crossing by the direct bonds or a up-down crossing by the dual bonds, so that without loss of generality the crossing probabilities are of order unity uniformly in $L$.  Standard arguments (see e.g. Theorem 2 in \cite{CS}) can then be used to demonstrate power law lower bounds.
\qed

\medskip
\noindent{\bf Remark.} Our assumptions of positive correlations and that $s_1=s_2=s_3$ are the ingredients needed to use the Zhang (and \cite{Russo}, \cite{GGR}) arguments.  Without these assumptions we cannot mathematically rule out the possibility of percolation \emph{before} or \emph{at} the self-dual point with unique infinite clusters of both types.  In the independent case, coexisting clusters were ruled out in \cite{GrimBook} using direct (Kesten--style) arguments.  It is conceivable that these arguments could be modified to the present case but we make no specific claims.  Nevertheless, some of the isotropic results can be extended outside the regions of positive correlations by domination arguments:

\begin{cor}
In the region $a >r_0/2$, $e<r_0/2$ the relevant (percolative) conclusions of Theorem \ref{PD} hold while in the region $a<r_0/2$, $e>r_0/2$ the relevant non-percolative conclusions of Theorem \ref{PD} hold.

\end{cor}
\noindent{\bf Proof: }Consider a point with parameters $a>r_0/2$, $e<r_0/2$ which is not covered in the previous theorem.  Such a point can be joined by a horizontal line to a point in the percolative region described in Theorem \ref{PD}.  For all intents and purposes, the new measure is obtained from the known percolative measure by replacing empty triangles with singly occupied triangles: Explicitly, the measure in question stochastically dominates a measure with the stated percolative properties.  The conclusion follows since the two claims about the regions $a>e$, $a+e>r_0$ may be phrased in terms of the events:\\
(1) The existence (wp1) of an infinite cluster and\\
(2) Uniqueness of said cluster.\\
The first is manifestly increasing while the second is equivalent to the absence of an infinite cluster of the dual type, hence also increasing.
The region $a<r_0/2$, $e>r_0/2$ is handled similarly.
\qed

\section{\large{Generalized Star--Triangle Relations -- Random Cluster Measure}}
\subsection{\large{Graphical Weights and Spin Systems}}
\label{phisiks}
We start in this subsection with the random cluster models -- a generalization of the usual random cluster models which features interactions among 
certain triples of sites.  Here we will confine attention only to 
triples which constitute three vertices of an up pointing triangle.  

The random-cluster models are defined by four parameters, $e$, $s$, $a$, 
and $q$, and are given formally by 
\begin{equation} 
\label{starr} 
 W(\omega) \varpropto q^{c(\omega)} s^{|{\sf s}(\omega)|}  a^{|{\sf a}(\omega)|}  e^{|{\sf e}(\omega)|}
 \end{equation}
where $\omega$ is a bond configuration, $|{\sf s}(\omega)|$ denotes the number of triangles with solely one side occupied and $|{\sf a}(\omega)|$ denotes the number of triangles with all three vertices connected, and $|{\sf e}(\omega)|$ the number of empty triangles.  It may be assumed, without loss of generality, that 
$a + 3s + e = 1$.
Of course as usual the above 
only makes sense in finite volumes with particular boundary conditions; 
infinite volume problems are extracted via limits.  However, as 
far as we are concerned, boundary conditions only enter through the 
definition of $c(\omega)$; once we establish the basic monotonicity properties of the model,  there are natural dominations in both volume and the various parameters $s$, $a$ and $e$.  
Then, the passage to infinite volume follows the 
exactly the same lines as for the usual random-cluster model.  Indeed, 
as far as these general matters are concerned we refer the reader to 
\cite{Grimmett} (see also \cite{Chayes} and \cite{CChayes}) where the issues have been discussed in some detail.

It is also clear (see the above mentioned citations) that for integer 
$q$ greater than one, this random-cluster model is the graphical 
representation of a (formal) Potts Hamiltonian with two and three site 
interactions:
\begin{equation} -\beta\mathcal{H} = \sum_{<x,y,z> }J 
(\delta_{\sigma_x\sigma_y} + \delta_{\sigma_y\sigma_z} + \delta_{\sigma_x\sigma_z}) + \kappa 
\delta_{\sigma_x\sigma_y\sigma_z}, \end{equation}
where the sum is over all generic up-pointing triangles.  We assume that $J$ is positive but there is, as of yet, no restriction concerning the parameter $\kappa$.

For completeness, a quick derivation proceeds as follows:
Let $\Lambda$ denote a finite collection of (up-pointing) triangles and $\mathcal{H}_\Lambda$ the restriction of $\mathcal{H}$ to $\Lambda$ with free boundary conditions, and $Z_{\Lambda}$ the corresponding partition function.  Then, 

\begin{equation*}  
Z_\Lambda = \sum_{\sigma_{\Lambda}}e^{-\beta \mathcal{H}_{\Lambda}} = \sum_{\sigma_{\Lambda}} \prod_{<x,y,z> \in \Lambda}  (S\delta_{\sigma_x\sigma_y} + 1)(S\delta_{\sigma_y\sigma_z} + 1)(S\delta_{\sigma_x\sigma_z} + 1) (1+g \delta{\sigma_x\sigma_y\sigma_z)},
\end{equation*}
where $S = e^{J} -1$ and $g=e^{\kappa}-1$, and again with no stipulation about the sign of $g$.  Multiplying everything out, we get
\begin{equation*} 
Z_\Lambda = \sum_{\sigma_{\Lambda}} \prod_{<x,y,z> \in \Lambda} [1 + S(\delta_{\sigma_x\sigma_y} + \delta_{\sigma_y\sigma_z} +\delta_{\sigma_x\sigma_z}) +A \delta_{\sigma_x\sigma_y\sigma_z}],
\end{equation*}where $A = 3S^2 + g (1 + S)^3$, which we \emph{now} stipulate to be positive.  Notice that we have deliberately failed to distinguish terms corresponding to products of two, versus three Kronecker deltas.
Opening up the product and identifying graphical terms in the usual fashion we perform the trace to obtain
\begin{equation}\label{partition}
Z_{\Lambda} = \sum_{\omega} q^{c(\omega)} S^{|{\sf s}(\omega)|} A^{|{\sf a}(\omega)|},
\end{equation}
where $\omega$ denotes a bond configuration restricted to five possibilities on each triangle as described in the previous section.  Since everything is positive, the objects in the above summand represent \emph{weights} for the configurations $\omega$.  For convenience, we can multiply the above by an overall (irrelevant) factor and then, by suitable redefinitions of parameters, we have our weights in the form of (\ref{starr}).

We remark that the more standard decomposition into eight configurations per triangle would, as can be checked, lead to positive correlations if and only if $g \geq 0$.  Indeed, $g/(1+g)$ corresponds  exactly to the parameter $t$ which was discussed in equations (\ref{32a})-(\ref{32c}).  As we will show in Theorem \ref{positive correlation} below, the present system provides a great deal more leeway.  

\medskip
\noindent
{\bf Remark.}  Finally, it is worth a reminder that as far as the spin systems are concerned, most quantity's relevance can be read directly from the graphical problem \cite{ACCN}, \cite{EdSo}.  In particular (at least in the realm of positive correlations), percolation is synonymous with magnetization, while the absence of percolation implies unicity among the possible limiting Gibbs states. 

\subsection{\large{Duality Relations and Self-Dual Curve}}
\begin{thm} For the random cluster measure as defined in the previous section, the duality relations are given by
\begin{equation*}\frac{s^*}{e^*} = \left(\frac{qs}{a} \right)\end{equation*}
and
\begin{equation*}\frac{a^*}{e^*} = \left(\frac{q^2e}{a} \right).\end{equation*}
The self dual curve, obtained in the above by setting $a=a^*$, $e=e^*$ and $s=s^*$ is then
\begin{equation*}a=qe.\end{equation*}
\end{thm}

\noindent{\bf Remark.} We note that the above corresponds exactly to equation (15) in \cite{Wu-Lin}.

\medskip

\noindent{\bf Proof (sketch): }To derive the duality relations, we make use of Euler's formula, which, as usual, has to be interpreted in the context of specific boundary conditions.  And here we have the additional step of integrating out the middle spin to return to the triangular lattice.  However, with careful consideration of the situation at the boundary, dual measures may be identified in finite volume.  Specifically, if $\Lambda$ consists of nothing more than $N$ connected up--pointing triangles with free boundary conditions, then the dual model will consist of the corresponding down--pointing triangles with fully wired boundary conditions.  Other scenarios at the boundary can be treated in a similar fashion;  we will be content to proceed formally.  But before we begin there is yet another technical difficulty: Our three-body interactions do not distinguish between triangles with two or three edges occupied; in order to use Euler's formula we must take this into account, so we set the convention that all three-body interactions have all three edges occupied.  Now, finally, we have: 
    \begin{equation*} \begin{split} W(\omega) &= q^{c(\omega)} s^{|{\sf s}(\omega)|} a^{|{\sf a}(\omega)|} e^{|{\sf e}(\omega)|}\\
    &\varpropto q^{l(\omega)} \left(\frac{s}{q}\right)^{|{\sf s}(\omega)|} \left(\frac{a}{q^3}\right)^{|{\sf a}(\omega)|}e^{|{\sf e}(\omega)|}.
    \end{split} \end{equation*}
 
\noindent Thus if $\omega^{\bigstar}$ is the standard dual (on the hexagonal lattice) we have:
\begin{equation*}
    W(\omega) \varpropto q^{c(\omega^\bigstar)} \left(\frac{s}{q}\right)^{|{\sf s}(\omega^\bigstar)|} \left(\frac{a}{q^3}\right)^{|{\sf e}(\omega^\bigstar)|}e^{|{\sf a}(\omega^\bigstar)|},
   \end{equation*}
  where $|e(\omega^\bigstar)|$ corresponds to the number of empty triads, etc.  Finally, integrating out all middle spins, we obtain:

\begin{equation*}\begin{split} W(\omega) &\varpropto q^{c(\omega^*)} \left(\frac{s}{q}\right)^{|{\sf s}(\omega^*)|} \left(\frac{a}{q^2}\right)^{|{\sf e}(\omega^*)|}e^{|{\sf a}(\omega^*)|}\\
   & \varpropto q^{c(\omega^*)} \left(\frac{qs}{a}\right)^{|{\sf s}(\omega^*)|}  1^{|{\sf e}(\omega^*)|}\left(\frac{eq^2}{a}\right)^{|{\sf a}(\omega^*)|}.
    \end{split} \end{equation*}
Here we have used the fact that the empty configuration on the triad has four connected components while that on the triangle when the middle vertex is integrated out has only three, so we must compensate a factor of $q$ for each ${\sf e}(\omega^\bigstar)$, yielding the $q^{-2}a$.  The weights are now in the form of equation (\ref{partition}).  Derivation of the self-dual curve is now straightforward.
\qed

Simple algebra now gives:
\begin{cor}\label {qinvar}For $\lambda \geq 0$, the regions $ae \geq \lambda s^2$ are invariant under the $*$--map.
\end{cor}

\subsection{Positive Correlation}
Our proof of positive correlations will concern $N$ triangles with configurations of the type described and measures determined by the weights given in (\ref{starr}).  For the purposes of this proof, we make no restrictions on the geometry of the triangles: they need not represent a subset of the triangular (or any other planar) lattice.  In general, sites can belong to any number of triangles, but if a pair of sites belong to two distinctive triangles, the associated bonds can appear twice.  In addition, we will need to consider different sorts of boundary conditions on our $N$ triangles; these will, generically, be denoted by $\Gamma$.  These $\Gamma$ conditions are the identification of sets of points which are considered to be ``preconnected'' (even if no bonds are present).  In particular, the specification of $\Gamma$ provides us with a precise notion of $c(\omega)$ and, for all intents and purposes, determines the geometry of the collection.  

\begin{defn}\label{measure}
Fix $a$, $e$, $s$ with $a+3s+e=1$.  Let $\Lambda$ be a fixed set of vertices in the triangular lattice corresponding to $M$ triangles which we label $t_1,t_2,\dots,t_M$.   Let $\mathcal{T}_M \equiv \{[\sfE], [\sfS]_1, [\sfS]_2, [\sfS]_3, [\sfA] \}$$^M$ denote the set of configurations on $\{t_1,t_2,\dots,t_M\}$.  Let $\Gamma$ denote an arbitrary wiring on $\Lambda$, then for $\omega \in \mathcal{T}_M$, 
\begin{equation}\label{weights}
W_\Lambda^\Gamma(\omega) \varpropto q^{c(\omega,\Gamma)}e^{|\sf{e}(\omega)|}s^{|\sf{s}(\omega)|}a^{|\sf{a}(\omega)|},
\end{equation}  
where $c(\omega,\Gamma)$ now denotes the number of connected components determined by the wiring $\Gamma$ as well as the configuration $\omega$.  Now for $N \leq M$, and $\omega \in \mathcal{T}_N$, we let 
\begin{equation*}
\mu_N^\Gamma(\omega) \varpropto 
W_\Lambda^\Gamma
(\omega,  \underset{ M-N \hspace{2mm} \text{times}}{\underbrace{[\sfE],\dots,[\sfE]}}),
\end{equation*}
denote the measure on those $N$ triangles obtained from the corresponding weight.
\end{defn}

\noindent{\bf Remark.}
The main thing to remember from the above definition is that we are working with some \emph{a priori} $\Lambda$ and \emph{all} the vertices of $\Lambda$ are taken into account in the term $c(\omega,\Gamma)$; this will become important later in the section when the structure of the weights actually come into play.  Needless to say, we will be interested (for the purposes of induction) in an $N$ which may be envisioned as far smaller than $M$; indeed, for finite $M$ there is no difficulty with the immediate passage $M \rightarrow \infty$.

\begin{thm}\label{positive correlation}
Let $\mu_N^\Gamma$ denote the measures as described above with $q\geq1$ and $ae \geq 2s^2$.  Then for all $N$ and all wiring boundary conditions $\Gamma$, these measures have positive correlations.
\end{thm}

The idea is to proceed by induction on the number of triangles $N$ which we regard as embedded in the larger space of $M$ triangles, $N-M$ of which are automatically empty.  We will need the strong inductive hypothesis that $\mu_{N-1}^\Gamma$ has positive correlations for \emph{all} possible wirings $\Gamma$. For the case $N=1$, there are clearly only five possible outside wirings: no vertices are connected, the vertices corresponding to side one (respctively two and three) are connected, and finally all three vertices are connected; we denote these wirings by $\mcE$, $\mcS_1$ $\mcS_2$, $\mcS_3$, and $\mcA$, respectively.  The all wired case, namely $\mu^{\mcA}_1$ is exactly the case proved in Theorem \ref{2t}.  Let us quickly dispense with another example, $\mu_1^{\mcS_1}$.  Here we see 
\begin{equation*} 
\mu_1^{\mcS_1} ([{\sf S}]_1)  = zs.
\end{equation*}
Meanwhile, for $k=2,3$, 
\begin{equation*}
 \mu_1^{\mcS_1} ([{\sf S}]_k)  = z\frac{s}{q},
 \end{equation*}
 and finally 
 \begin{equation*}
 \mu_1^{\mcS_1 }([{\sf A}])  = z\frac{a}{q},  \hspace{2mm} \mu_1^{\mcS_1} ([{\sf E}])  = ze,
 \end{equation*}
 where we use the notation $[\sfA]$, $[\sfS]_1$, $\dots$,$[\sfE]$ to denote the relevant corresponding events and $z$ is a normalization constant.  The necessary inequality $\mu_1^{\mcS_1}([\sfA]) \mu_1^{\mcS_1}([\sfE]) \geq \mu_1^{\mcS_1}([\sfS]_1) (\mu_1^{\mcS_1}([\sfS]_2) + \mu_1^{\mcS_1}([\sfS]_3)$ follows readily from $ae \geq 2s^2$ provided $q \geq 1$.  The other cases are just as easily demonstrated and we may consider the base case to be established.

We make use of two key ideas in the forthcoming inductive proof.  The first is a generalized version of the lattice condition.  Indeed, whenever the underlying space is the product of linearly ordered spaces, the lattice condition is entirely equivalent to the minimalist version:
\begin{equation}
\label{1P2}
\frac{\nu(\eta, a, b)}{\nu(\eta, a, b^{\prime})} \geq \frac{\nu(\eta, a^{\prime}, b)}{\nu(\eta, a^{\prime}, b^{\prime})}, \end{equation}where the $a$'s and $b$'s represent the configuration at any two coordinates, $\eta$ is all other coordinates and $a \geq a^{\prime}$ and $b \geq b^{\prime}$.  Crucial to our argument is that despite the absence (or inapplicability) of the full lattice condition, an analogue of (\ref{1P2}) nevertheless holds.  The second key idea is a slight generalization of Proposition 2.22 in \cite{Li} which is the statement that a convex combination of two measures with positive correlations itself has positive correlations if one of the measures FKG dominates the other.  We state and prove these as our next two lemmas below.

\begin{lemma} \label{1P2T} Let $\mu_N^\Gamma$ be defined as above with $q\geq1$.  Then an analogue of (\ref{1P2}) holds for $\mu_N^\Gamma$, provided the separate increases pertain to different triangles.  E.g., if $\mbbT_{N-2}$ is the configuration on the first $N-2$ triangles, and we have $T_{N-1}$, $T_{N-1}^\prime$, $T_N$, $T_N^\prime$ as configurations on the last two triangles with $T_{N-1}\succeq T_{N-1}^\prime$ and $T_N\succeq T_N^{\prime}$, then
\begin{equation*}
\frac{\mu_N^\Gamma(\mbbT_{N-2}, T_{N-1}, T_N)}{\mu_N^\Gamma(\mbbT_{N-2}, T_{N-1}, T_N^\prime)} \geq \frac{\mu_N^\Gamma(\mbbT_{N-2}, T_{N-1}^\prime, T_N)}{\mu_N^\Gamma(\mbbT_{N-2}, T_{N-1}^{\prime}, T_N^\prime)}.
\end{equation*}
\end{lemma}
\noindent{\bf Proof: }Examining the ratios in the statement above, a quick glance at (\ref{weights}) reveals that all the ``prefactors'' cancel on both sides of the purported inequality, leaving only the $q$--dependent terms.  Since $q>1$, the above amounts to a special case of 
\begin{equation*}
C(\omega \vee \eta) + C(\omega \wedge \eta) \geq C(\omega) + C(\eta),
\end{equation*} 
which has been proved in complete generality in numerous places (e.g. \cite{ACCN}).
\qed

\begin{lemma} \label{lll} Let $(\mcH, \succeq_{_{\mcH}})$ and $(\mcK, \succeq_{_{\mcK}})$ be finite partially ordered sets.  Let $\mu$ be a probability measure on $\mcH$ and for each $\eta \in \mcH$, let $\nu_\eta$ be a probability measure on $\mcK$.  It is supposed that $\mu$ has positive correlations, that for each $\eta$, the measure $\nu_{\eta}$ has positive correlations and furthermore, if $\eta_1  \succeq \eta_2$, then $\nu_{\eta_1} \underset{\text{\tiny{FKG}}} \geq \nu_{\eta_2}$.  Then 
\begin{equation*} 
\nu(-) \equiv \sum_{\eta \in \mcH} \mu(\eta) \nu_\eta (-) 
\end{equation*} 
has positive correlations.
\end{lemma}
\noindent{\bf Proof: }Let $f$ and $g$ be increasing functions on $\mcK$.  Then
\begin{equation*}\begin{split}\mbfE_{\nu}(fg) &= \sum_{\omega \in \mcK} \nu(\omega)f(\omega)g(\omega)\\
&= \sum_{\eta \in \mcH} \mu(\eta)\mbfE_{\nu_\eta}(fg)\\
&\geq \sum_{\eta \in \mcH} \mu(\eta)\mbfE_{\nu_\eta}(f)\mbfE_{\nu_\eta}(g).\end{split}
\end{equation*}
It is observed from the hypothesis that $\mbfE_{\nu_\eta}(f)$ and $\mbfE_{\nu_\eta}(g)$ are increasing in $\eta$ and the result follows from the positive correlation of $\mu$.
\qed

\medskip
Now let us informally proceed with an inductive proof.  In what is to follow we assume that$f$ and $g$ are increasing functions on $N$ triangles, $\mbbT_{N-1}$ always denotes the configuration on the first $N-1$ triangles and $T_N \in \{[\sfA], \dots, [\sfE] \}$ a generic state of the $N^{th}$ triangle.  We condition on the state of the last triangle, and according to Bayes' formula, we get
\begin{equation*}
\mu_N^\Gamma(-) = \sum_{T_N} \mu^\Gamma_{N|\Delta_N} (T_N) \mu_N^\Gamma(-|T_N),
\end{equation*}  
where $\mu^\Gamma_{N|\Delta_N}$ is the restriction of $\mu^\Gamma_N$ to the last triangle.

As far as the first $N-1$ triangles are concerned, we can apply the inductive hypothesis to conclude that the measures $\mu^{\Gamma}_N(-|T_N)$ has positive correlations, since the conditioning, along with $\Gamma$, give us \emph{some} wiring scenario for these triangles.  So  (appealing to Lemma (\ref{lll})) we will be done if we can show that (i) $\mbfE_N^{\Gamma}(f|T_N)$ and $\mbfE_N^{\Gamma}(g|T_N)$ are increasing in $T_N$ (i.e. $\mu_N^\Gamma(-|T_N) \underset{\text{\tiny{FKG}}}\geq \mu_N^\Gamma(-|T_N^\prime)$ whenever $T_N \succeq T_N^\prime$), and (ii) the measure $\mu^{\Gamma} _{N|\Delta_N}$  has positive correlation.  These are the topics of yet the next two lemmas.

\begin{lemma}\label{increasing} Let $f$ and $T_{N}$ be as described and define 
\begin{equation*} F_{T_N} = \mbfE_N^\Gamma(f|T_N). \end{equation*}
Then $F_{T_N}$ is an increasing function. \end{lemma}
  
\noindent{\bf Proof: }Suppose $T_N \succeq T_N^\prime$.  Then we note that, as in the standard argument, Lemma (\ref{1P2T}) implies that
\begin{equation}\label{*} 
\phi(\mbbT_{N-1}) = \frac{\mu_N^\Gamma(T_N^\prime)}{\mu_N^\Gamma(\mbbT_{N-1},T_N^\prime)} \frac{\mu_N^\Gamma(\mbbT_{N-1}, T_N)}{\mu_N^\Gamma(T_N)} 
\end{equation}
is an increasing function of $\mbbT_{N-1} = (T_1, ..., T_{N-1})$.  
We aim to show that 
\begin{equation*} 
\mbfE_N^\Gamma(f|T_N) \geq \mbfE_N^\Gamma(f|T_N^\prime). 
\end{equation*}We have\begin{equation*}\begin{split} \mbfE_N^\Gamma(f|T_N) &= \sum_{\mbbT_{N-1}} f(\mbbT_{N-1}, T_N) \frac{\mu_N(\mbbT_{N-1}, T_N)}{\mu_N(T_N)}\\
&\geq \sum_{\mbbT_{N-1}} f(\mbbT_{N-1}, T_N^\prime) \frac{\mu_N(\mbbT_{N-1}, T_N)}{\mu_N(T_N)},\end{split}\end{equation*}
since $f$ is increasing and $T_N \geq T_N^\prime$.  Now the last expression can be rewritten as 
\begin{equation*}\sum_{\mbbT_{N-1}}f(\mbbT_{N-1},T_N^\prime)  \frac{\mu_N^\Gamma(\mbbT_{N-1},T_N^\prime)}{\mu_N^\Gamma(T_N^\prime)} \phi(\mbbT_{N-1}) = \mbfE_N^\Gamma(f\phi|T_N^\prime),\end{equation*}
which by induction is greater than or equal to $\mbfE_N^\Gamma(f|T_N^\prime)\mbfE_N^\Gamma(\phi|T_N^\prime)$.  Thus, concatenating the above expressions, we have
\begin{equation*}\begin{split}\mbfE_N^\Gamma(f|T_N) & \geq \mbfE_N^\Gamma(f|T_N^\prime)\mbfE_N^\Gamma(\phi|T_N^\prime)\\
&= (\sum_{\mbbT_{N-1}} f(\mbbT_{N-1},T_N^\prime) \frac{\mu_N^\Gamma(\mbbT_{N-1}, T_N^\prime)}{\mu_N^\Gamma(T_{N-1}^\prime)})(\sum_{\mbbT_{N-1}} \frac{\mu_N^\Gamma(\mbbT_{N-1}, T_N)}{\mu_N^\Gamma(T_N)})\\
&= \mbfE_N^\Gamma(f|T_N^\prime),\end{split}\end{equation*}
since$\sum_{\mbbT_{N-1}} \frac{\mu_N^\Gamma(\mbbT_{N-1}, T_N)}{\mu_N^\Gamma(T_N)}=1$.\qed

\begin{lemma}\label{poscor}
Let $\mu^\Gamma_{N| \Delta_N}$ denote the measure $\mu_N^\Gamma$ as described above restricted to the $N^{th}$ triangle.  Then $\mu^\Gamma_{N| \Delta_N}$ has positive correlation.
\end{lemma}
\noindent{\bf Proof: }We will again make use of Lemma \ref{lll}, so we write
\begin{equation}\label{bayes} \mu^\Gamma_{N| \Delta_N}(-) = \sum_{\mcB} \mu_N^\Gamma(\mcB) \mu^{\mcB}_{N|\Delta_{N}}(-), \end{equation}
where $\mcB$ represents the total wiring conditions outside the $N^{th}$ triangle due to the initial wiring condition $\Gamma$ \emph{and} the outside configurations, $\mbbT_{N-1}$.  However, the overall effect of $\Gamma$ and $\mbbT_{N-1}$ is to produce one of the five types of wiring on a single triangle -- a situation with which we are familiar -- and henceforth we may assume $\mcB \in \{\mcA, \mcS_1, \mcS_2, \mcS_3, \mcE\}$. 

We note that for each $\mcB$,  $\mu^{\mcB}_{\Delta_N}$ has positive correlations (the subject of the base case).  Next, we claim that $\mcB \succeq \mcB^\prime$ implies that $\mu_N^{\mcB} \underset{\text{\tiny{FKG}}}\geq \mu_N^{\mcB^\prime}$. This follows from the observation that less wiring on the outside produces more factors of $q^{-1}$ for the weights of the higher configurations (see the Remark after Definition (\ref{measure})).  Explicitly, it can be checked that for $\mcB \succeq \mcB^\prime$,
\begin{equation*}
\mu_N^{\mcB^\prime} \varpropto D\mu_N^{\mcB},
\end{equation*}
where $D$ (which depends on $\mcB^\prime$ and $\mcB$) is a decreasing function of $T_N$.  Thus we have verified two of the three hypotheses of Lemma (\ref{lll}).  

We are down to the last hypothesis; here we will need to write $\mu_N^\Gamma(\mcB)$ in a more explicit form.  Note that by induction $\mu_{N-1}^\Gamma(\mcB)$ has positive correlations, so we seek some relationship between $\mu_N^\Gamma(\mcB)$ and $\mu_{N-1}^\Gamma(\mcB)$.  To do this we must exploit the ``almost'' product structure of the weights (\ref{weights}) from which our measures came.  So we first let $Z_{N-1}({\mcB})$ denote the \emph{weight} of observing $\mcB$, before the introduction of the $N^{th}$ triangle, and let $Z_{N-1}^T = \sum_{\mcB} Z_{N-1}(\mcB)$ denote the overall normalization factor, so that $\lambda_{\mcB} \equiv Z_{N-1}(\mcB)/Z_{N-1}^T = \mu_{N-1}^\Gamma(\mcB)$.  Next we may write  
$Z_{N-1}(\mcB)=\sum_{\mbbT_{N-1}}\mathbf{1}_{\mbbT_{N-1}\cup\Gamma=\mcB}W_\Lambda^{\Gamma}(\mbbT_{N-1})$,
where $W_\Lambda^{\Gamma}(\mbbT_{N-1})$ is the weight of observing the configuration $\mbbT_{N-1}$ as given by (\ref{weights}).
Similarly, if $Z_{N}({\mcB})$ denotes the weight of observing $\mcB$ \emph{given} the $N^{th}$ triangle, then
$Z_{N}(\mcB)=\sum_{\mbbT_{N-1}}\mathbf{1}_{\mbbT_{N-1}\cup\Gamma=\mcB}\left(\sum_{T_N}W_\Lambda^{\Gamma}(\mbbT_{N-1},T_N)\right).$
Comparing the previous two expressions and referring back to Definition (\ref{measure}), it is not difficult to see that
\begin{equation*}
Z_{N}(\mcB)=n_{_{\mcB}}Z_{N-1}(\mcB), 
\end{equation*} 
where (up to a factors of $e$) $n_{_{\mcE}}=(\frac{a}{q^2} + \frac{3s}{q} + e)$, $n_{_{\mcS_i}}= (\frac{a}{q} + s + \frac{2s}{q} + e)$, and $n_{_{\mcA}}=(a + 3s + e)$ -- which happens to be one.  Thus, letting $Z_N^T=\sum_{\mcB}Z_N(\mcB)$, we arrive at
\begin{equation*}
\mu_N^\Gamma(\mcB) = \frac{Z_N(\mcB)}{Z_N^T} = \frac{Z_{N-1}^T}{Z_N^T} \lambda_{\mcB}n_{_{\mcB}}.
\end{equation*}  
It is noted that the factor $Z^T_{N-1}/Z^T_N$ is independent of the wiring $\mcB$, $T_N$, etc.  Thus by Theorem (\ref{2t}) all we need to show is that $(n_{_{\mcA}}\lambda_{\mcA})(n_{_{\mcE}}\lambda_{\mcE})$ exceeds $n_{_{\mcS_1}}\lambda_{\mcS_1}(n_{_{\mcS_2}}\lambda_{\mcS_2}+n_{_{\mcS_3}}\lambda_{\mcS_3})$ -- or whatever ordering combination maximizes the latter object. To this end, let $\sigma$ be a permutation on three letters such that 
$\lambda_{\mcS_{\sigma_1}} \geq \lambda_{\mcS_{\sigma_2}}$ and  $\lambda_{\mcS_{\sigma_1}} \geq \lambda_{\mcS_{\sigma_3}}$. Our last hypothesis will be verified if we can show that
\begin{equation*}
(n_{_{\mcE}}\lambda_{\mcE})(n_{_{\mcA}}\lambda_{\mcA})\geq  (n_{_{\mcS_{\sigma_1}}}\lambda_{\mcS_{\sigma_1}})(n_{_{\mcS_{\sigma_2}}}\lambda_{\mcS_{\sigma_2}} + n_{_{\mcS_{\sigma_3}}}\lambda_{\mcS_{\sigma_3}}).
\end{equation*}
To this end, we first observe that the induction hypothesis implies $\lambda_{\mcA}\lambda_{\mcE} \geq \lambda_{\mcS_{\tau(1)}}(\lambda_{\mcS_{\tau(2)}} + \lambda_{\mcS_{\tau(3)}}$) for any permutation on three letters $\tau$:
On general grounds this is true because of the similarity between the outside wiring space and the inside configuration space.  But, to proceed formally, let $f$ and $g$ be the increasing functions of the outside wiring such that $f(\mcE)=g(\mcE)=0$, $f(\mcA)=g(\mcA)=1$, $f(\mcS_{\tau(1)})=1-g(\mcS_{\tau(1)}) = 1$ and $f(\mcS_{\tau(i)})=1-g(\mcS_{\tau(i)})=0$, $i=2,3$.  Then by the fact that $\mu_{N-1}$ has positive correlation, we indeed get $\lambda_{\mcA}\lambda_{\mcE} \geq \lambda_{\mcS_{\tau(1)}}(\lambda_{\mcS_{\tau(2)}} + \lambda_{\mcS_{\tau(3)}})$.
On the basis of this inequality we only need that $n_{_{\mcA}}n_{_{\mcE}} \geq n_{_{\mcS_{\sigma_1}}}^2$ (since $n_{_{\mcS_{\sigma_1}}}n_{_{\mcS_{\sigma_1}}} = n_{_{\mcS_{\sigma_1}}}n_{_{\mcS_{\sigma_2}}}=
n_{_{\mcS_{\sigma_1}}}n_{_{\mcS_{\sigma_3}}}$), i.e. we need that, 
\begin{equation*}
\label{nlambda}
(a + 3s + e)(\frac{a}{q^2} + \frac{3s}{q} + e) \geq (\frac{a}{q} + s + \frac{2s}{q} + e)^2. 
\end{equation*}
Now if one multiplies and compares terms, one has an expression which involves $(q-1)$ times a quantity which is ``easily positive''.

We have verified all three hypotheses of Lemma (\ref{lll}) and can therefore conclude that $\mu^\Gamma_{N| \Delta_N}$ has positive correlation. \qed 

\bigskip
\noindent {\bf Proof of Theorem (\ref {positive correlation}):}  As already remarked, we will (again) use Lemma \ref{lll}.  Explicitly, we apply Lemma (\ref{lll}) with $\mcH=\{[\sfA], [\sfS]_1, \dots,[\sfE]\}$ (corresponding to configurations on the $N^{th}$ triangle) and $\mu=\mu^\Gamma_{N|_{\Delta_N}}$, and $\mcK=\{[\sfA], [\sfS]_1, \dots,[\sfE]\}$$^N$ (corresponding to configurations on all $N$ triangles) and $\nu_\eta = \mu_N^\Gamma(-|\eta)$.  The three hypotheses of the Lemma are verified by the induction hypothesis and Lemmas \ref{increasing} and \ref{poscor}.\qed

\bigskip
We conclude this section with a brief discussion on infinite-volume limits: In the region of positive correlations, more wiring leads to a higher measure.  Thus, for free boundary conditions (the restrictions of) measures increase with increase volume and for fully wired boundary conditions, they decrease.  So, for a nested sequence of volumes which exhaust the lattice, well-defined infinite-volume limits -- which are independent of sequence -- exist.  Furthermore, as mentioned earlier, wired and free measures may be dually identified in finite volume.  Thus, in turn, we may identify the dual of the infinite volume free measure as the wired measure and vice versa.  

\subsection{Phase Transitions}
In this subsection, we establish results on phase transitions in the $q$--state Potts/random cluster models under consideration.  Here, unlike in the percolation case, we cannot establish with certainty whether the transition is continuous or discontinuous.  Moreover, for the continuous cases, our statements will be considerably weaker than our Theorem \ref{PD} since much of the technical artillery (e.g. the van den Berg-Kesten inequalities) do not apply.  In particular, in the continuous case, we cannot even prove that the percolation/magnetization transition actually occurs on the self-dual line.  Nevertheless, critical \emph{behavior} is established for self--dual points which are also points of continuity, the subject of our first proposition:

\begin{prop}Consider the random cluster model on the triangular lattice as defined by (\ref{starr}) and satisfying $ae \geq 2s^2$.  Then at any self--dual point $a=qe$ which is a point of continuity of the bond density the following hold: (1) The percolation probability vanishes and (2) there are power law lower bounds on the correlation functions.
\end{prop}
\noindent{\bf Proof: }Much of the proof can be transcribed directly from our proof of Theorem \ref{PD} and as for the rest, similar arguments have appeared before (\cite{BC}, \cite{CS}), so we will be succinct.  The first statement follows from the results in \cite{GKR} which, under the conditions of positive correlations and 2D symmetries, forbids coexisting infinite clusters of the opposite types.  Thus, in any realization, there is either no percolation of either type or there are separate states (depending on how the infinite--volume limit was constructed) which have and don't have percolation.  However, this latter happenstance, by appeal to Strassen's Theorem \cite{Strassen} implies that the distinctive states have different bond densities which would imply a discontinuity in the bond density.  For the second statement, routine arguments which may be traced back to \cite{ACCN} imply that the limiting random cluster measure is unique and therefore may be identified with the dual measure; on this basis the rest of the argument follows mutatis mutantis from the proof of Theorem \ref{PD} for percolation (again see \cite{CS}).\qed

\medskip
Finally we show that in the region of positive correlation, any discontinuity in bond density must occur on the self--dual curve:
\begin{prop}Consider the random cluster model on the triangular lattice as defined by (\ref{starr}) and satisfying $ae \geq 2s^2$.  Then any discontinuity in the bond density must occur on the self--dual curve as given by $a=qe$.
\end{prop}
\noindent{\bf Proof: }Our proof is a variation of the one found in \cite{BC}; here we unfortunately do not have a convenient family of curves which are nicely preserved under the duality relations.  We will work with $A$ and $S$ parameters given in (\ref{partition}); suppose at $(A_0, S_0)$ -- with $A_0>2S_0^2$ -- we have a discontinuity in the bond density.  Let $\lambda \geq 1$ and consider the curve $C \equiv \{(2\lambda S_0^2, S_0): \lambda \geq 1\}$.  We note that along $C$ the measure is FKG increasing with increasing $\lambda$.  Next let $\lambda^{SD}$, $\lambda^P$ and $\lambda^D$ denote the corresponding values of $\lambda$ at which the curve $C$ intersects the self--dual curve, the percolation threshold, and the discontinuity, respectively.  We aim to show that $\lambda^{SD}=\lambda^P=\lambda^D$.

Let $C_l$ denote the part of $C$ which is below the self--dual curve and similarly let $C_u$ denote the part of $C$ which is above the self--dual curve.  Since it is not the case that $(C_l)^* = C_u$, we need to define two new curves to work with: Let $C_y = C_l \cup (C_l)^*$ and $C_p=C_u \cup (C_u)^*$, and we parametrize $C_y$ by $\lambda_y$ and $C_p$ by $\lambda_p$ with the requirement that $\lambda=\lambda_y$ on $C_l$ and $\lambda = \lambda_p$ on $C_u$ (and extending in the obvious fashion).  We remark that with these parametrizations, $C_y$ and $C_p$ are FKG increasing in $\lambda_y$ and $\lambda_p$ by duality (or by explicit computation): E.g., on $C_y \cap C_l$ the measures are clearly FKG increasing; on the other hand, this implies the measures corresponding to the \emph{dual} parameters -- these lie on $(C_l)^*$ and are parametrized by $\lambda_y \geq \lambda_y^{SD}$ -- are \emph{decreasing} in $\lambda_y$ for $\lambda_y \leq \lambda_y^{SD}$, and hence increasing in $\lambda_y$ for $\lambda_y \geq \lambda_y^{SD}$.  Now if we let $\lambda_y^D$, $\lambda_y^{SD}$ and $\lambda_y^P$ denote the corresponding values of $\lambda_y$ at which the curve $C$ intersects the self--dual curve, the percolation threshold and (should it exist) the discontinuity, respectively.  Similarly we define $\lambda_p^D$, $\lambda_p^{SD}$ and $\lambda_p^{P}$ for $C_p$.  Then $\lambda_y^{SD}=\lambda_p^{SD}=\lambda^{SD}$. 

First we show that $\lambda^P \geq \lambda^{SD}$: If this is not the case, then for $\lambda^{SD} > \lambda > \lambda^P$ the direct model is percolating in the wired state.  Note that this $\lambda$ corresponds to a $\lambda_y$ in our new parametrization.  At the dual value, $\lambda_y^*$, we would then have dual percolation in the state with free boundary conditions.  However, the dual model in the wired state at parameter $\lambda$ ``dominates'' the dual model in the free state at parameter $\lambda_y^*$, and hence there is dual \emph{and} direct percolation at $\lambda$ (e.g. in the wired state), which is a contradiction of \cite{GKR}.  Next we can easily show that $\lambda^P \leq \lambda^D$: This is because a discontinuity in the bond density implies non-uniqueness of the limiting measure and hence, ultimately, percolation.  Finally, we must have $\lambda^D \leq \lambda^{SD}$: Towards a contradiction assume that $\lambda^{D} > \lambda^{SD}$; this implies that $\lambda_p^D$ actually exists and is equal to $\lambda^D$.  Next note that the same argument that showed $\lambda^P \leq \lambda^D$ also shows $\lambda_p^P \leq \lambda_p^D$.  Since we have a discontinuity in the direct model if and only if we have a discontinuity in the dual model, we have another discontinuity at $\lambda_p^* < \lambda_p^{SD} \leq \lambda_p^P$, a contradiction.
\qed

\section{Conclusion}
We have described a Potts/random cluster model on the triangular lattice with three--body correlations.  By introducing a reduced state space, the duality relations are easily derived.  It is noted, in the context of spin systems, that the purely ferromagnetic region of parameters is \emph{not} mapped into itself under duality.  More generally, in the $q\geq1$ random cluster models, when we consider the full state space, the region which has positive correlations is not mapped into itself.  However, for the reduced case, necessary and sufficient conditions for positive correlations are derived which are invariant under duality and include a larger portion of the original parameter space.  Under the conditions of positive correlations, for percolation and for values of $q$ where there are discontinuities, it is proved that the transition occurs at the self--dual point; if there is no discontinuity, self--dual points admit critical behavior.  On the basis of exact solutions \cite{BaxterTA} it has been argued that the dividing line is $q=4$, similar to the situation on the square lattice.  The advantage of the current random cluster formulation is that this hypothesis can be tested numerically using cluster methods; e.g., the algorithms in \cite{SW}, \cite{MCLSC} and \cite{CM2} can be readily adapted.  While we have no reason to doubt the results in \cite{BaxterTA} in this case, for a related model with three--body interactions on the square lattice, there is some evidence pointing to the reduction of the dividing $q$.  In any case, although we will not discuss details, it should at least be possible to prove that for large $q$ there is a discontinuous transition.  Here certain modifications will be needed to adapt the methods of reflection positivity to the present case, which may very well be the subject of a later paper.

\section*{Acknowledgments}
\noindent 
This research was supported by the NSF grant~DMS-0306167.

\end{document}